\documentclass[12pt]{iopart}

\usepackage{iopams}
\usepackage{graphicx}

\newcommand{\ket}[1]{\left|#1\right\rangle}
\newcommand{\bra}[1]{\left\langle#1\right|}

\begin{document}
	
	\title[Single-Step Hybrid CV–DV Transfer of Multipartite W States Using Cat-State Qubits]{Single-Step Hybrid CV–DV Transfer of Multipartite W States Using Cat-State Qubits}
	
	\author{Muhammad Nehal Khan$^{1}$, Sumrah Hussain$^{1}$}
	
	\address{$^1$Department of Physics, School of Computing, Data Science and Physics, William and Mary, Williamsburg, Virginia 23185, USA}
	\ead{mkhan05@wm.edu}
	\vspace{10pt}
	\begin{indented}
		\item[]January 2026
	\end{indented}

	\begin{abstract}
		We propose a deterministic hybrid continuous-variable - discrete-variable (CV-DV) scheme
		for the single-step transfer of an $n$-qubit W state encoded in photonic Schr$\ddot{o}
		$dinger
		cat-state qubits within a circuit QED architecture. Logical qubits are encoded in even- and
		odd-parity cat states of bosonic modes, while effective Raman-type interactions between
		resonator pairs are mediated by a single superconducting flux qutrit operating in the
		dispersive regime. The protocol coherently transfers the multipartite W state in a single
		collective operation without populating higher excited atomic levels, thereby strongly
		suppressing decoherence. Numerical simulations based on the full Lindblad master equation,
		including realistic cavity dissipation, qutrit relaxation and dephasing, and inter-cavity
		crosstalk, show that a three-qubit cat-state W state can be transferred with a maximum
		fidelity of approximately $0.92$. These results demonstrate the feasibility of scalable
		hybrid CV-DV entanglement transfer using current circuit QED technology.
	\end{abstract}

	%\maketitle
	\section{Introduction}
	
	Cavity quantum electrodynamics (QED) provides a powerful platform for quantum information processing through the coherent interaction of quantized electromagnetic fields with atomic or artificial atomic systems confined in high-quality resonators \cite{1,2,3}. In particular, cavity and circuit-based QED architectures have enabled scalable quantum networks, long-distance quantum communication, and multipartite entanglement generation \cite{4,5,6,7,8,9,10,11,12}. Among multipartite entangled states, W states play a central role due to their robustness against particle loss and their importance in quantum communication protocols \cite{13}.
	
	In the solid-state domain, circuit QED based on superconducting qubits interacting with microwave resonators has emerged as one of the leading candidates for large-scale quantum information processing \cite{14,15,16,17,18,19,20,21,22}. This architecture has enabled high-fidelity quantum gates \cite{23,24,25,26,27,28,29,30}, quantum algorithms \cite{31}, quantum memories \cite{32,33,35,36}, and the controlled generation of multipartite entangled states \cite{37,38,39,40,41,42,43}. In particular, several schemes have been proposed and experimentally demonstrated for generating W states of superconducting qubits via dispersive or resonant interactions \cite{44,45,46,47,48,49, 68}.
	
	More recently, attention has shifted toward hybrid continuous-variable-discrete-variable (CV-DV) encodings, where logical qubits are encoded in nonclassical states of bosonic modes. A prominent example is the cat-state qubit (cqubit), where logical states are represented by even and odd-parity Schr$\ddot{o}
$dinger cat states of a harmonic oscillator. Cat-state qubits are intrinsically resilient against single-photon loss and can be stabilized and protected using quantum error correction techniques \cite{50,51}. Significant progress has been made toward realizing cat-state entanglement \cite{52}, single and two-qubit gates \cite{53,54,55,56}, and multi-qubit controlled operations \cite{57}. Furthermore, experimental demonstrations of cat-state transfer between resonators have been reported \cite{58,59}.
	
	Despite these advances, the direct transfer of multipartite W states encoded in cat-state qubits across spatially separated cavities has not yet been thoroughly explored. In particular, scalable schemes that operate in a single step, avoid population of higher excited atomic levels, and remain compatible with current circuit QED technology remain an open challenge. In this work, we address this gap by proposing a scheme for transferring an $n$-qubit W state encoded in cat states between two sets of $n$ microwave resonators. The protocol employs $2n$ resonators coupled to a single superconducting flux qutrit, enabling effective cavity-cavity interactions while suppressing unwanted crosstalk through frequency selectivity. The protocol is deterministic, requires no measurement, and operates entirely within a hybrid CV-DV framework.
	
	\section{Transfer of W state encoded with cat-states}
	We consider a hybrid continuous-variable-discrete-variable (CV-DV) architecture consisting of $2N$ bosonic microwave resonators coupled to a single superconducting three-level artificial atom (qutrit), as illustrated in Fig.~1(a). Each pair of resonators $\{2j-1,2j\}$ encodes one logical qubit using orthogonal Schr$\ddot{o}
$dinger cat states, thereby realizing a DV qubit within a CV Hilbert space. The superconducting qutrit, with energy eigenstates $\ket{g}$, $\ket{e}$, and $\ket{f}$ [Fig.~1(b)], acts as a common quantum bus that mediates effective interactions between the resonator pairs.
	
	To suppress unwanted inter-resonator crosstalk, each resonator is assumed to be strongly detuned from all non-targeted qutrit transitions. Specifically, within each resonator pair $\{2j-1,2j\}$, the odd-indexed resonator couples dispersively to the $\ket{g}\leftrightarrow\ket{f}$ transition, while the even-indexed resonator couples dispersively to the $\ket{e}\leftrightarrow\ket{f}$ transition. The corresponding detunings are chosen such that $|\Delta_{(2j-1)(2j)}|\gg g_{2j-1},g_{2j}$, ensuring that the upper level $\ket{f}$ remains only virtually populated throughout the dynamics. In addition, a classical microwave pulse is applied resonantly to the $\ket{g}\leftrightarrow\ket{e}$ transition of the qutrit, while remaining far detuned from all other transitions. This drive enables controlled phase evolution without inducing population transfer to the upper level $\ket{f}$. Experimentally, both the qutrit transition frequencies and the resonator frequencies can be tuned on nanosecond timescales, allowing the required dispersive conditions to be satisfied dynamically\cite{60,61}.
	
	\begin{figure}\label{fig1}
		\centering
		\includegraphics[width=16cm, height=6cm]{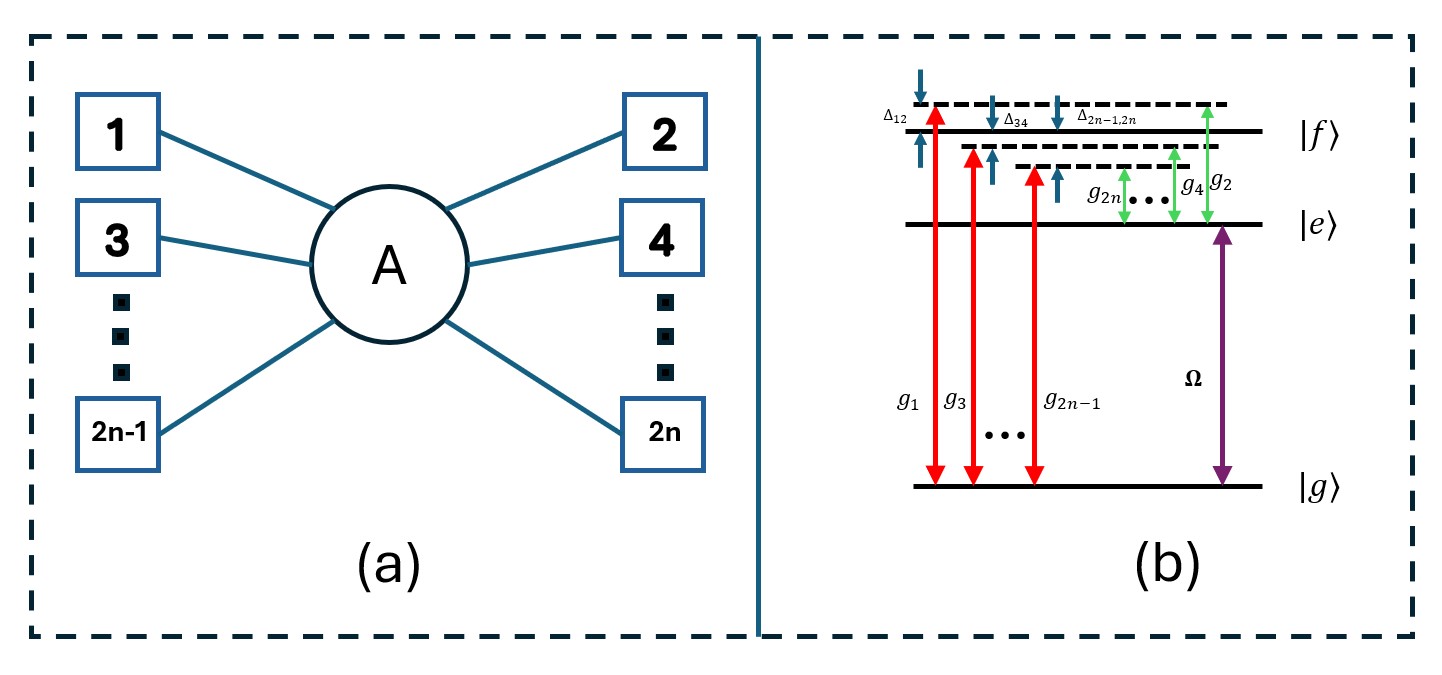}
		\caption{(a) Diagram of an artificial superconducting qutrit (represented by circle A) and six microwave resonators. For 1D microwave resonator, the qutrit is capacitively coupled to each resonator \cite{59}. (b) Illustration of the qutrit-cavity dispersive interaction and the qutrit-pulse resonant interaction.}
	\end{figure}
	
	Using the above conditions, and by applying the rotating wave approximation, we get the Hamiltonian in interaction picture as below (assuming $\hbar=1$)
	%	\begin{eqnarray}\label{eq1}
	%		H =& (g_1 e^{i \Delta_{12} t} \hat{a}_1 \sigma^+_{fg} + g_2 e^{i \Delta_{12} t} \hat{a}_2 \sigma^+_{fe}) 
	%		+ (g_3 e^{i \Delta_{34} t} \hat{a}_3 \sigma^+_{fg} + g_4 e^{i \Delta_{34} t} \hat{a}_4 \sigma^+_{fe}) \nonumber \\
	%		& + (g_5 e^{i \Delta_{56} t} \hat{a}_5 \sigma^+_{fg} + g_6 e^{i \Delta_{56} t} \hat{a}_6 \sigma^+_{fe}) + \Omega \sigma_{eg}^+ + H.c.,
	%	\end{eqnarray}
	%	
	%	
	\begin{eqnarray}\label{eq1}
		H =& (g_1 e^{i \Delta_{12} t} \hat{a}_1 \sigma^+_{fg}
		+ g_2 e^{i \Delta_{12} t} \hat{a}_2 \sigma^+_{fe}) \nonumber\\
		&+ \sum_{j=2}^{N}
		\left(
		g_{2j-1} e^{i \Delta_{2j-1,\,2j} t} \hat{a}_{2j-1} \sigma^+_{fg}
		+ g_{2j}   e^{i \Delta_{2j-1,\,2j} t} \hat{a}_{2j}   \sigma^+_{fe}
		\right) \nonumber\\
		&+ \Omega \sigma^+_{eg} + \mathrm{H.c.}
	\end{eqnarray}
	Here, $\sigma^{+}_{fg}=\ket{f}\bra{g}$, $\sigma^{+}_{fe}=\ket{f}\bra{e}$, and $\sigma^{+}_{eg}=\ket{e}\bra{g}$ are the raising operators of the qutrit, $\Omega$ denotes the Rabi frequency of the applied microwave pulse, and $\hat{a}_j$ ($\hat{a}_j^\dagger$) is the annihilation (creation) operator of the $j$th resonator. Equation~(\ref{eq1}) describes the hybrid CV-DV interaction in the dispersive regime, which forms the basis for engineering effective beam-splitter-type couplings between resonator pairs.

	In the large-detuning regime, $|\Delta_{(2j-1)(2j)}|\gg g_{2j-1},g_{2j}$, the upper level $\ket{f}$ of the qutrit is only virtually excited and can be adiabatically eliminated. As a result, effective Raman-type interactions are induced between the $\ket{g}$ and $\ket{e}$ states via each resonator pair $\{2j-1,2j\}$. By choosing sufficiently different detunings for distinct
	resonator pairs, unwanted cross-pair Raman processes are strongly suppressed, allowing each logical qubit to evolve independently. Under these conditions, the effective Hamiltonian takes the form given in Eq.~(\ref{eq2})~\cite{62}.

	%	\begin{eqnarray}\label{eq2}
	%		%\fl
	%		H_{eff} = -2\lambda_1 \hat{a}^{\dagger}_1 \hat{a}_1 \sigma_{gg} -2\lambda_2 \hat{a}^{\dagger}_2 \hat{a}_2 \sigma_{ee} - 2\lambda_3 \hat{a}^{\dagger}_3 \hat{a}_3 \sigma_{gg} \nonumber \\
	%		-2\lambda_4 \hat{a}^{\dagger}_4 \hat{a}_4 \sigma_{ee} - 2\lambda_5 \hat{a}^{\dagger}_5 \hat{a}_5 \sigma_{gg} -2\lambda_6 \hat{a}^{\dagger}_6 \hat{a}_6 \sigma_{ee} \nonumber \\
	%		-2\lambda_{12} (\hat{a}_1 \hat{a}^{\dagger}_2 \sigma_{eg}^+ + \hat{a}^{\dagger}_1 \hat{a}_2 \sigma_{eg}^-)-2\lambda_{34} (\hat{a}_3 \hat{a}^{\dagger}_4 \sigma_{eg}^+ + \hat{a}^{\dagger}_3 \hat{a}_4 \sigma_{eg}^-) \nonumber \\
	%		-2\lambda_{56}(\hat{a}_5 \hat{a}^{\dagger}_6 \sigma_{eg}^+ + \hat{a}^{\dagger}_5 \hat{a}_6 \sigma_{eg}^-) +\Omega \sigma_{x}
	%	\end{eqnarray}

	\begin{eqnarray}\label{eq2}
		H_{eff} =& -2\lambda_1 \hat{a}^{\dagger}_1 \hat{a}_1 \sigma_{gg}
		-2\lambda_2 \hat{a}^{\dagger}_2 \hat{a}_2 \sigma_{ee} \nonumber\\
		&- \sum_{j=2}^{N}\Big(
		2\lambda_{2j-1}\hat{a}^{\dagger}_{2j-1}\hat{a}_{2j-1}\sigma_{gg}
		+ 2\lambda_{2j}\hat{a}^{\dagger}_{2j}\hat{a}_{2j}\sigma_{ee}
		\Big) \nonumber\\
		&-2\lambda_{12}\Big(\hat{a}_1 \hat{a}^{\dagger}_2 \sigma_{eg}^+
		+\hat{a}^{\dagger}_1 \hat{a}_2 \sigma_{eg}^- \Big) \nonumber\\
		&- \sum_{j=2}^{N} 2\lambda_{(2j-1)(2j)}\Big(
		\hat{a}_{2j-1}\hat{a}^{\dagger}_{2j}\sigma_{eg}^+
		+ \hat{a}^{\dagger}_{2j-1}\hat{a}_{2j}\sigma_{eg}^- \Big)
		+ \Omega \sigma_{x}.
	\end{eqnarray}

	where $\sigma^{-}_{eg} = \ket{g}\bra{e}$, $\sigma_{gg} = \ket{g}\bra{g}$, $\sigma_{ee} = \ket{e}\bra{e}$, $\sigma_{x} = \sigma^{+}_{eg} + \sigma^{-}_{eg}$, $\lambda_1 = g^2_1/(2\Delta_{12})$, $\lambda_2 = g^2_2/(2\Delta_{12})$, $\lambda_3 = g^2_3/(2\Delta_{34})$, $\lambda_4 = g^2_4/(2\Delta_{34})$, $\lambda_5 = g^2_5/(2\Delta_{56})$, $\lambda_6 = g^2_6/(2\Delta_{56})$, $\lambda_{12} = g_1 g_2/(2\Delta_{12})$, $\lambda_{34} = g_3 g_4/(2\Delta_{34})$, and $\lambda_{56} = g_5 g_6/(2\Delta_{56})$.
	
	The first four terms of Eq.~(\ref{eq2}) describe photon-number-dependent ac-Stark shifts of the qutrit energy levels induced by each resonator mode. The remaining terms correspond to effective Raman-mediated beam-splitter interactions within each resonator pair $\{2j-1,2j\}$, conditioned on the qutrit state. These interactions form the physical mechanism enabling coherent transfer of
	hybrid-encoded logical qubits.

	By preparing the qutrit in the dressed state $\ket{+}=(\ket{g}+\ket{e})/\sqrt{2}$, and operating in the strong-driving regime $2\Omega \gg \lambda_{2j-1},\lambda_{2j},\lambda_{(2j-1)(2j)}$, rapidly oscillating terms can be neglected under the rotating-wave
	approximation. Consequently, the qutrit remains effectively decoupled from the dynamics, and the system evolution is governed solely by pairwise beam-splitter Hamiltonians acting on the bosonic modes, as shown in Eq.~(\ref{eq6}). This decoupling is essential for realizing deterministic hybrid CV-DV W-state transfer without populating 	the excited qutrit level.

	By introducing the dressed-state basis $\ket{\pm} = \frac{1}{\sqrt{2}}(\ket{g} \pm \ket{e})$, the operators associated with the superconducting three-level system appearing in Eq.~(\ref{eq2}) can be expressed in terms of the dressed-state operators as $\sigma_{gg} = \frac{1}{2}\left(I + \sigma^{+} + \sigma^{-}\right), \sigma_{ee} = \frac{1}{2}\left(I - \sigma^{+} - \sigma^{-}\right), \sigma_{eg}^{+} = \frac{1}{2}\left(\sigma_{z} + \sigma^{+} - \sigma^{-}\right), \sigma_{eg}^{-} = \frac{1}{2}\left(\sigma_{z} - \sigma^{+} + \sigma^{-}\right)$, where the dressed-state Pauli operators are defined as $\sigma_{z} = \ket{+}\bra{+} - \ket{-}\bra{-}, \qquad \sigma^{+} = \ket{+}\bra{-}, \qquad \sigma^{-} = \ket{-}\bra{+}$. In this representation, the transverse operator satisfies $\sigma_{x}=\sigma_{z}$.
	
	Using the above transformations, Eq.~(\ref{eq2}) can be rewritten in the dressed-state basis, where the Hamiltonian contains rapidly oscillating terms proportional to $e^{\pm i 2\Omega t}$, with $\Omega$ denoting the Rabi frequency of the classical driving field. In the strong-driving regime, defined by $ 2\Omega \gg \lambda_j,\; \lambda_{(2j-1)(2j)} \qquad (j=1,2,\dots)$,
	all terms oscillating with such high frequencies and may be neglected by implementing the rotating-wave approximation. Thus, we obtained the following Hamiltonian

	\begin{eqnarray}\label{eq3}
		H_{eff} =&
		-\left(\lambda_1 \hat{a}_1^\dagger \hat{a}_1
		+\lambda_2 \hat{a}_2^\dagger \hat{a}_2 \right) \nonumber\\
		&- \sum_{j=2}^{N}
		\left(
		\lambda_{2j-1}\hat{a}_{2j-1}^\dagger \hat{a}_{2j-1}
		+\lambda_{2j}\hat{a}_{2j}^\dagger \hat{a}_{2j}
		\right) \nonumber\\
		&- \lambda_{12}
		\left(\hat{a}_1 \hat{a}_2^\dagger
		+\hat{a}_1^\dagger \hat{a}_2\right)\sigma_z \nonumber\\
		&- \sum_{j=2}^{N}
		\lambda_{(2j-1)(2j)}
		\left(
		\hat{a}_{2j-1}\hat{a}_{2j}^\dagger
		+\hat{a}_{2j-1}^\dagger\hat{a}_{2j}
		\right)
		+ \Omega \sigma_z .
	\end{eqnarray}

	Applying unitary transformation on $e^{-i 2 \Omega t}$, with $H_0 = -\lambda_1 \hat{a}_1^\dagger \hat{a}_1
	+\lambda_2 \hat{a}_2^\dagger \hat{a}_2 - \sum_{j=2}^{N} \lambda_{2j-1}\hat{a}_{2j-1}^\dagger \hat{a}_{2j-1}
	+\lambda_{2j}\hat{a}_{2j}^\dagger \hat{a}_{2j} + \Omega \sigma_z$, we obtain 
	%	\begin{eqnarray}\label{eq4}
	%		H_e = e^{i H_0 t} (H_{eff}-H_0) e^{-i H_0 t} \nonumber \\
	%		= -\lambda(\hat{a}_1 \hat{a}^\dagger_2 + \hat{a}^\dagger_1 \hat{a}_2) \sigma_z +\lambda(\hat{a}_3 \hat{a}^\dagger_4 + \hat{a}^\dagger_3 \hat{a}_4) \sigma_z + \lambda(\hat{a}_5 \hat{a}^\dagger_6 + \hat{a}^\dagger_5 \hat{a}_6) \sigma_z
	%	\end{eqnarray}	
	
	\begin{eqnarray}\label{eq4}
		H_e &=& e^{i H_0 t} (H_{eff}-H_0) e^{-i H_0 t} \nonumber\\
		&=& -\lambda\left(\hat{a}_1 \hat{a}^\dagger_2 + \hat{a}^\dagger_1 \hat{a}_2\right)\sigma_z
		+ \sum_{j=2}^{N} \lambda\left(\hat{a}_{2j-1}\hat{a}^\dagger_{2j}
		+ \hat{a}^\dagger_{2j-1}\hat{a}_{2j}\right)\sigma_z ,
	\end{eqnarray}
	
	%	where we can set the following conditions
	%	\begin{eqnarray}\label{eq5}
	%		\lambda_1 = \lambda_2, \, \lambda_3 = \lambda_4, \, \lambda_5 = \lambda_6
	%	\end{eqnarray}
	%	\begin{eqnarray}\label{eq5b}
	%		\lambda = \lambda_{12} = -\lambda_{34} = -\lambda_{56}
	%	\end{eqnarray}
	
	\begin{eqnarray}\label{eq5}
		\lambda_1=\lambda_2, \qquad
		\lambda_{2j-1}=\lambda_{2j}\ \ (j=2,3,\ldots,N).
	\end{eqnarray}

	We can easily prepare the qutrit in $\ket{+}$ state by applying the $\pi$-pulse to the transition level $\ket{g} \leftrightarrow \ket{e}$ resonantly, which is initially in the $\ket{g}$ state. The state $\ket{+}$ will have not any impact of the Hamiltonian (4), therefore qutrit will remain in that state. Thus, we can neglect the qutrit part and the effective Hamiltonian (4) will further simplified to
	%	\begin{eqnarray}\label{eq6}
	%		H_e = H_{e1} + H_{e2} + H_{e3}
	%	\end{eqnarray}	
	%	with
	%	\begin{eqnarray}\label{eq7}
	%		H_{e1} = -\lambda(\hat{a}^\dagger_1 \hat{a}_2 + \hat{a}_1 \hat{a}^\dagger_2)
	%	\end{eqnarray}
	%	\begin{eqnarray}\label{eq8}
	%		H_{e2} = \lambda(\hat{a}^\dagger_3 \hat{a}_4 + \hat{a}_3 \hat{a}^\dagger_4)
	%	\end{eqnarray}	
	%	\begin{eqnarray}\label{eq9}
	%		H_{e3} = \lambda(\hat{a}^\dagger_5 \hat{a}_6 + \hat{a}_5 \hat{a}^\dagger_6)
	%	\end{eqnarray}	
	
	\begin{eqnarray}\label{eq6}
		H_e = H_{e1} + \sum_{j=2}^{N} H_{ej}
	\end{eqnarray}
	with
	\begin{eqnarray}\label{eq7}
		H_{e1} = -\lambda\left(\hat{a}^\dagger_1 \hat{a}_2
		+ \hat{a}_1 \hat{a}^\dagger_2\right)
	\end{eqnarray}
	\begin{eqnarray}\label{eq8}
		H_{ej} = \lambda\left(\hat{a}^\dagger_{2j-1}\hat{a}_{2j}
		+ \hat{a}_{2j-1}\hat{a}^\dagger_{2j}\right),
		\qquad j=2,3,\ldots,N .
	\end{eqnarray}
	
	The Hamiltonian presented in Eq.(\ref{eq6}) describes independent qutrit-mediated beam-splitter interactions within each resonator pair $(2j-1,2j)$ ($j=1,2,\ldots,N$). These interactions act exclusively on the bosonic modes and preserve the photon-number parity in each resonator, thereby protecting the logical qubit encoding realized in the cat-state basis. Consequently, this Hamiltonian enables the deterministic transfer of a hybrid continuous-variable-discrete-variable (CV-DV) $n$-qubit W state from the set of odd-indexed resonators $\{1,3,\ldots,2N-1\}$ to the corresponding set of even-indexed resonators $\{2,4,\ldots,2N\}$.

	Initially, the logical $n$-qubit W state is encoded in the odd-indexed resonators $\{1,3,\ldots,2N-1\}$ using cat-state qubits, while all even-indexed resonators $\{2,4,\ldots,2N\}$ are prepared in the vacuum state. In this hybrid CV-DV encoding, each logical qubit is realized by a pair of bosonic modes, and the W state consists of a single odd-parity cat state coherently and symmetrically delocalized among the $n$ logical qubits, with all remaining modes occupying even-parity cat states.

	To encode logical qubits within the bosonic resonators, we employ Schr$\ddot{o}
$odinger cat states of definite photon-number parity. In this hybrid CV-DV encoding, the even-parity cat state represents the logical state $\ket{0}$, while the odd-parity cat state represents the logical state $\ket{1}$. These states form an orthogonal basis that is robust against single-photon loss and naturally compatible with circuit QED platforms~\cite{50,51}.
	
	%	\begin{eqnarray}\label{eq10}
	%		\ket{cat} = \Sigma^{\infty}_{m=0} C_{2m} \ket{2m}, \,\,\,\,\, \ket{\overline{cat}} = \Sigma^{\infty}_{n=0} C_{2n+1} \ket{2n+1} 
	%	\end{eqnarray}

	\begin{eqnarray}\label{eq10}
		\ket{\mathrm{cat}} &=& \sum_{m=0}^{\infty} C_{2m}\ket{2m}, \nonumber\\
		\ket{\overline{\mathrm{cat}}} &=& \sum_{n=0}^{\infty} C_{2n+1}\ket{2n+1}.
	\end{eqnarray}
	where $n$ and $m$ are non-negative integers, $C_{2m} = 2 N^+_{\alpha} e^{- |\alpha|^2/2} \alpha^{2m} / \sqrt{(2m)!}$, and $C_{2n+1} = 2 N^-_{\alpha} e^{- |\alpha|^2/2} \alpha^{2n+1} / \sqrt{(2n+1)!}$. We can noticed that $\ket{cat}$ state is orthogonal to $\ket{\overline{cat}}$ state, independent of $\alpha$ (except for $\alpha = 0$). The cat state $\ket{cat}$ is representing $\ket{0}$ logical state while the other cat-state $\ket{\overline{cat}}$ is representing $\ket{1}$ logical state in quantum optics.
	
	The transfer of the n-qubit W state from odd-numbered resonators to the even-numbered resonators is described by
	%	\begin{eqnarray}\label{eq11}
	%		%\fl
	%		\frac{1}{\sqrt{3}} (\ket{cat}_1\ket{cat}_3\ket{\overline{cat}}_5 + \ket{cat}_1\ket{\overline{cat}}_3\ket{cat}_5 + \ket{\overline{cat}}_1\ket{cat}_3\ket{cat}_5) \ket{0}_2 \ket{0}_4\ket{0}_6 \nonumber \\ 
	%		%\fl
	%		\rightarrow \ket{0}_1 \ket{0}_3\ket{0}_5 \frac{1}{\sqrt{3}} (\ket{cat}_2\ket{cat}_4\ket{\overline{cat}}_6 + \ket{cat}_2\ket{\overline{cat}}_4\ket{cat}_6 + \ket{\overline{cat}}_2\ket{cat}_4\ket{cat}_6)
	%	\end{eqnarray}
	
	\begin{equation}\label{eq11}
		\ket{W_n^{(\mathrm{cat})}}
		=\frac{1}{\sqrt{n}}\sum_{j=1}^{n}
		\ket{\mathrm{cat}}_1\ket{\mathrm{cat}}_2\cdots
		\ket{\overline{\mathrm{cat}}}_j
		\cdots\ket{\mathrm{cat}}_n .
	\end{equation}

	According to the cat-state definitions given in Eq.~(\ref{eq10}), the Fock states of the $j$th resonator can be written as $\ket{2m}_j=(\hat{a}_j^\dagger)^{2m}\ket{0}_j/\sqrt{(2m)!}$ and $\ket{2n+1}_j=(\hat{a}_j^\dagger)^{2n+1}\ket{0}_j/\sqrt{(2n+1)!}$. Using these relations, the even- and odd-parity Schr$\ddot{o}
$dinger cat states that encode a logical qubit in the hybrid CV-DV architecture can be expressed, for an arbitrary resonator $j\,(j=1,2,\ldots,2N)$, as

	\begin{eqnarray}\label{eq12}
		\ket{cat}_j = \Sigma^{\infty}_{m=0} C^{\prime}_{2m} (\hat{a}^{\dagger}_j)^{2m} \ket{0}_j, \,\,\,\,\,
		\overline{\ket{cat}}_j = \Sigma^{\infty}_{n=0} C^{\prime}_{2n+1} (\hat{a}^{\dagger}_j)^{2n+1} \ket{0}_j,
	\end{eqnarray}
	
	where, $C^{\prime}_{2m} = C_{2m}/\sqrt{(2m)!}$ and $C^{\prime}_{2n+1} = C_{2n+1}/\sqrt{(2n+1)!}$.
	
	From Eq.~(\ref{eq6}), the system dynamics is governed by the effective Hamiltonian $H_e=\sum_{j=1}^{N} H_{ej}$, where each $H_{ej}$ describes an independent beam-splitter-type interaction within the resonator pair $\{2j-1,2j\}$. Since these pairwise Hamiltonians commute with one another, i.e., \[ [H_{ej},H_{e\ell}]=0, \qquad j\neq \ell, \] the global time-evolution operator factorizes into a product of independent two-mode evolutions. Consequently, the evolution of an $N$-qubit W state encoded in hybrid CV-DV cat-state qubits can be evaluated by considering the action of each $H_{ej}$ separately, leading to the following state evolution.

\begin{eqnarray}\label{eq13}
	\!\!\!\!
	e^{-iH_et}
	\Bigg[
	\Big(\prod_{j=1,\, j\neq k}^{N}
	\ket{\mathrm{cat}}_{2j-1}\Big)
	\ket{\overline{\mathrm{cat}}}_{2k-1}
	\Big(\prod_{j=1}^{N}\ket{0}_{2j}\Big)
	\Bigg]
	\nonumber\\[6pt]
	=
	\prod_{j=1}^{N}
	\left(
	e^{-iH_{ej}t}\ket{\psi_j}_{2j-1}\ket{0}_{2j}
	\right)
	\nonumber\\[6pt]
	=
	\prod_{j=1,\, j\neq k}^{N}
	\Bigg[
	\sum_{m=0}^{\infty} C'_{2m}
	\left(
	e^{-iH_{ej}t}\hat a_{2j-1}^\dagger e^{iH_{ej}t}
	\right)^{2m}
	\ket{0}_{2j-1}\ket{0}_{2j}
	\Bigg]
	\nonumber\\[6pt]
	\otimes
	\Bigg[
	\sum_{n=0}^{\infty} C'_{2n+1}
	\left(
	e^{-iH_{ek}t}\hat a_{2k-1}^\dagger e^{iH_{ek}t}
	\right)^{2n+1}
	\ket{0}_{2k-1}\ket{0}_{2k}
	\Bigg].
\end{eqnarray}

	where we have used
	\begin{eqnarray}
		e^{-iH_{ej}t}\ket{0}_{2j-1}\ket{0}_{2j}=\ket{0}_{2j-1}\ket{0}_{2j},
		\qquad j=1,2,\ldots,N,
	\end{eqnarray}
	since each $H_{ej}$ annihilates the two-mode vacuum state.
	
	%	Note that $e^{-iH_{e1}t}\hat{a}^\dagger_1 e^{iH_{e1}t} = \cos(\lambda t) \hat{a}^{\dagger}_1+i\sin(\lambda t)\hat{a}^\dagger_2$, $e^{-iH_{e2}t}\hat{a}^\dagger_3 e^{iH_{e2}t} = \cos(\lambda t) \hat{a}^{\dagger}_3-i\sin(\lambda t)\hat{a}^\dagger_4$, and $e^{-iH_{e3}t}\hat{a}^\dagger_5 e^{iH_{e3}t} = \cos(\lambda t) \hat{a}^{\dagger}_5-i\sin(\lambda t)\hat{a}^\dagger_6$. For $\lambda t = \pi/2$, we have $e^{-iH_{e1}t}\hat{a}^\dagger_1 e^{iH_{e1}t} = i\hat{a}^\dagger_2$, $e^{-iH_{e2}t}\hat{a}^\dagger_3 e^{iH_{e2}t} = -i\hat{a}^\dagger_4$, and $e^{-iH_{e3}t}\hat{a}^\dagger_5 e^{iH_{e3}t} = -i\hat{a}^\dagger_6$. Thus, for $t = T = \pi/(2\lambda)$, we have from Eqs. (13) and (14)
	\begin{eqnarray}
		e^{-iH_{e1}t}\hat{a}^\dagger_{1}e^{iH_{e1}t}
		&=& \cos(\lambda t)\,\hat{a}^\dagger_{1}
		+i\sin(\lambda t)\,\hat{a}^\dagger_{2}, \nonumber\\
		e^{-iH_{ej}t}\hat{a}^\dagger_{2j-1}e^{iH_{ej}t}
		&=& \cos(\lambda t)\,\hat{a}^\dagger_{2j-1}
		-i\sin(\lambda t)\,\hat{a}^\dagger_{2j}, \qquad j=2,3,\ldots,N.
	\end{eqnarray}
	For $\lambda t=\pi/2$, we have
	\begin{eqnarray}
		e^{-iH_{e1}t}\hat{a}^\dagger_{1}e^{iH_{e1}t}
		&=& i\hat{a}^\dagger_{2}, \nonumber\\
		e^{-iH_{ej}t}\hat{a}^\dagger_{2j-1}e^{iH_{ej}t}
		&=& -i\hat{a}^\dagger_{2j}, \qquad j=2,3,\ldots,N.
	\end{eqnarray}
	Thus, for $t=T=\pi/(2\lambda)$, we have from Eqs.~(13) and (14)

\begin{eqnarray}\label{eq16gen}
	e^{-iH_eT}\Bigg[
	\Big(\prod_{\begin{array}{c} j=1\\ j\neq k \end{array}}^{N}
	\ket{\mathrm{cat}}_{2j-1}\Big)\,
	\ket{\overline{\mathrm{cat}}}_{2k-1}\,
	\Big(\prod_{j=1}^{N}\ket{0}_{2j}\Big)
	\Bigg]
	=
	\nonumber\\[6pt]
	\Big(\prod_{j=1}^{N}\ket{0}_{2j-1}\Big)\otimes
	\Bigg[
	\prod_{\begin{array}{c} j=1\\ j\neq k \end{array}}^{N}
	\left(
	\sum_{m=0}^{\infty}
	C'_{2m}\,e^{-im\pi}\,\ket{2m}_{2j}
	\right)
	\Bigg]
	\nonumber\\[6pt]
	\otimes
	\Bigg[
	\sum_{n=0}^{\infty}
	C'_{2n+1}\,e^{-i(2n+1)\pi/2}\,\ket{2n+1}_{2k}
	\Bigg].
\end{eqnarray}
		where we have used
	\[
	\ket{2m}_j=\frac{(\hat a_j^\dagger)^{2m}}{\sqrt{(2m)!}}\ket{0}_j,
	\qquad
	\ket{2n+1}_j=\frac{(\hat a_j^\dagger)^{2n+1}}{\sqrt{(2n+1)!}}\ket{0}_j,
	\quad (j=2,4,6),
	\]
	together with the definitions of the coefficients $C'_{2m}$ and $C'_{2n+1}$ given above.
	
	Equation (\ref{eq16gen}) shows that, at the interaction time
	$T=\pi/(2\lambda)$, each logical cat-state qubit is deterministically
	transferred from resonator $2j-1$ to resonator $2j$, while preserving
	the global $N$-qubit W-state superposition.

	After returning to the original interaction picture, the time evolution for the initial state of the whole system is given by

	\begin{eqnarray}
		\ket{\mathrm{cat}(\theta_{2j})}_{2j} &=& \sum_{m=0}^{\infty} C_{2m}\,e^{i\,2m\,\theta_{2j}}\ket{2m}_{2j},\\
		\ket{\overline{\mathrm{cat}}(\theta_{2j})}_{2j} &=& \sum_{n=0}^{\infty} C_{2n+1}\,e^{i\,(2n+1)\,\theta_{2j}}\ket{2n+1}_{2j}.
	\end{eqnarray}
	
	\begin{eqnarray}\label{eq19gen}
		e^{-iH_{0}\tau}e^{-iH_{e}\tau}\;
		\frac{1}{\sqrt{N}}
		\sum_{k=1}^{N}
		\Bigg[
		\ket{\overline{\mathrm{cat}}}_{2k-1}
		\prod_{\begin{array}{c} j=1\\ j\neq k \end{array}}^{N}
		\ket{\mathrm{cat}}_{2j-1}
		\Bigg]
		\times
		\Bigg(\prod_{j=1}^{N}\ket{0}_{2j}\Bigg)\ket{+}=
		\nonumber\\[8pt]
		e^{-i\phi_0}\;
		\Bigg(\prod_{j=1}^{N}\ket{0}_{2j-1}\Bigg)
		\times
		\frac{1}{\sqrt{N}}
		\sum_{k=1}^{N}
		\Bigg[
		\ket{\overline{\mathrm{cat}}(\theta_{2k})}_{2k}
		\prod_{\begin{array}{c} j=1\\ j\neq k \end{array}}^{N}
		\ket{\mathrm{cat}(\theta_{2j})}_{2j}
		\Bigg]\ket{+}.
	\end{eqnarray}
	
	%	where from line 1 to lines 2 and 3 we have used the results given in Eqs. (15) and (16). Here, $\phi_0 = \Omega \pi /(2\lambda)$, $\eta_{12} = \lambda_2/(2\lambda) + 1/2$, $\eta_{34} = \lambda_4/(2\lambda) - 1/2$, and $\eta_{56} = \lambda_6/(2\lambda) - 1/2$. We set
	%%\begin{eqnarray}\label{eq20}
	%		\lambda_2 = \lambda
	%	\end{eqnarray}
	%	\begin{eqnarray}\label{eq21}
	%		\lambda_4 = \lambda_6 = -\lambda
	%	\end{eqnarray}
	
Equation (\ref{eq19gen}) explicitly shows that the hybrid
CV-DV encoded $N$-qubit W state is coherently transferred
from the set of resonators $\{2j-1\}$ to $\{2j\}$ in a single
deterministic step, while the superconducting qutrit remains
factorized in the state $\ket{+}$ throughout the evolution.

	Here, $\phi_0=\Omega\pi/(2\lambda)$. We further define
	\begin{eqnarray}\label{eq:eta12}
		\eta_{12}=\frac{\lambda_2}{2\lambda}+\frac{1}{2},
	\end{eqnarray}
	and, for the remaining mode pairs,
	\begin{eqnarray}\label{eq:etaj}
		\eta_{(2j-1)(2j)}=\frac{\lambda_{2j}}{2\lambda}-\frac{1}{2}, \qquad j=2,3,\ldots,N.
	\end{eqnarray}
	same above and below
	\begin{eqnarray}\label{eq20}
		\lambda_2=\lambda,
	\end{eqnarray}
	\begin{eqnarray}\label{eq21}
		\lambda_{2j}=-\lambda, \qquad j=2,3,\ldots,N,
	\end{eqnarray}
	
	%	which leads to $\eta_{12} = 1$ and $\eta_{34} = \eta_{56} = -1$. For $\eta = 1$ and $\eta_{34} = \eta_{56} = -1$, we have $exp(i2\eta_{12} m \pi) = exp(i2\eta_{34} m\pi) = exp(i2\eta_{56} = 1$ and $exp(i2\eta_{12} (2n+1) \pi) = exp(i2\eta_{34}(2n+1)\pi) = exp(i2\eta_{34}(2n+1)\pi) = -1$. Thus, the state of the six cavities, given in Eq. (\ref{eq19}), becomes
	%	
	which leads to
	\[
	\eta_{12}=1, \qquad
	\eta_{(2j-1)(2j)}=-1,\quad j=2,3,\ldots,N .
	\]
	For these values of the parameters, the phase factors simplify as
	\begin{eqnarray}
		e^{i2\eta_{(2j-1)(2j)} m\pi} &=& 1, \qquad j=1,2,\ldots,N, \nonumber\\
		e^{i2\eta_{(2j-1)(2j)} (2n+1)\pi} &=& -1, \qquad j=1,2,\ldots,N .
	\end{eqnarray}
	Thus, all even-parity components acquire a trivial phase, whereas all odd-parity components pick up a relative minus sign.

	%%	\begin{eqnarray}\label{eq22}
	%%	%	\fl
	%%		\ket{0}_1 \ket{0}_3 \ket{0}_5 \frac{1}{\sqrt{3}} (\Sigma^{\infty}_{m=0} C_{2m} \ket{2m}_2 \otimes \Sigma^{\infty}_{m=0} C_{2m} \ket{2m}_4 \otimes \Sigma^{\infty}_{n=0} C_{2n+1} \ket{2n+1}_6 \nonumber\\ 
	%%		+ \Sigma^{\infty}_{m=0} C_{2m} \ket{2m}_2 \otimes \Sigma^{\infty}_{n=0} C_{2n+1} \ket{2n+1}_4 \otimes \Sigma^{\infty}_{m=0} C_{2m} \ket{2m}_6 \nonumber\\
	%%		+ \Sigma^{\infty}_{n=0} C_{2n+1} \ket{2n+1}_2 \otimes \Sigma^{\infty}_{m=0} C_{2m} \ket{2m}_4 \otimes \Sigma^{\infty}_{m=0} C_{2m} \ket{2m}_4)
	%%	\end{eqnarray}
	\begin{eqnarray}\label{eq22gen}
		\!\!\!\!\!\!\!\!\!\!\!\!\!\!\!\!\!\!\!\!\!\!\!\!\!\!\!\!\!\!
		\Bigg(\prod_{j=1}^{N}\ket{0}_{2j-1}\Bigg)\,
		\frac{1}{\sqrt{N}}
		\sum_{k=1}^{N}
		\Bigg[
		\left(\prod_{\begin{array}{c} j=1\\ j\neq k \end{array}}^{N}
		\sum_{m=0}^{\infty} C_{2m}\ket{2m}_{2j}\right)
		\otimes
		\left(\sum_{n=0}^{\infty} C_{2n+1}\ket{2n+1}_{2k}\right)
		\Bigg].
	\end{eqnarray}
	
	where the common phase factor $e^{-i\phi_0}$ has been omitted. According to Eq. (\ref{eq13}), the state (\ref{eq22gen}) can be written as
	%	\begin{eqnarray}\label{eq23}
	%	%	\fl
	%		\ket{0}_1 \ket{0}_3 \ket{0}_5 \frac{1}{\sqrt{3}} (\ket{cat}_2 \ket{cat}_4 \ket{\overline{cat}}_6 + \ket{cat}_2 \ket{\overline{cat}}_4 \ket{cat}_6 + \ket{\overline{cat}}_2 \ket{cat}_4 \ket{cat}_6),
	%	\end{eqnarray}
	\begin{eqnarray}\label{eq23gen}
		\Bigg(\prod_{j=1}^{N}\ket{0}_{2j-1}\Bigg)\,
		\frac{1}{\sqrt{N}}
		\sum_{k=1}^{N}
		\left[
		\ket{\overline{\mathrm{cat}}}_{2k}
		\prod_{\begin{array}{c} j=1\\ j\neq k \end{array}}^{N}
		\ket{\mathrm{cat}}_{2j}
		\right].
	\end{eqnarray}
	
The above expression demonstrates the deterministic transfer of a hybrid
continuous-variable-discrete-variable (CV-DV) encoded $N$-qubit W state from
the set of odd-indexed resonators $\{1,3,\ldots,2N-1\}$ to the corresponding
even-indexed resonators $\{2,4,\ldots,2N\}$. In this protocol, each logical qubit
is encoded in a pair of bosonic modes using orthogonal Schr$\ddot{o}
$dinger cat states,
and the W-state coherence is preserved throughout the transfer process.

The transfer operation relies on precise control of the qutrit energy-level
spacings to ensure that the qutrit remains effectively decoupled from all
resonator modes once the desired interaction time is reached
\cite{56}. Alternatively, the same decoupling condition can be achieved by
dynamically tuning the resonator frequencies, which is routinely accessible in
circuit QED architectures \cite{57}. These control techniques guarantee that
the effective interaction is switched off after the single-step transfer,
thereby freezing the transferred hybrid W state.

The result presented in Eq.~(\ref{eq24gen}) follows from the parameter conditions
specified in Eqs.~(\ref{eq5}), (\ref{eq21}), and their $N$-mode generalizations.
In particular, the equal-coupling conditions
$g_{2j-1}=g_{2j}$ $(j=1,2,\ldots,N)$ ensure symmetric Raman-mediated interactions
within each resonator pair. Such coupling requirements can be satisfied in both
one-dimensional and three-dimensional resonator implementations. For 3D
cavities, the coupling strengths $g_j$ can be engineered through appropriate
design of the qutrit loop geometry and cavity field profiles, while in 1D
transmission-line resonators they can be tuned via the capacitances $C_j$
between each resonator and the superconducting qutrit.

	We can easily confimrs that all the conditions from Eq. (\ref{eq6}) and Eq. (\ref{eq22gen}) will lead to
	%	\begin{eqnarray}\label{eq24}
	%		g_1 g_2/\Delta_{12} = -g_3 g_4/\Delta_{34} = -g_5 g_6/\Delta_{56}
	%	\end{eqnarray}
	\begin{eqnarray}\label{eq24gen}
		\frac{g_1 g_2}{\Delta_{12}}
		= -\,\frac{g_{2j-1}g_{2j}}{\Delta_{(2j-1)(2j)}},
		\qquad j=2,3,\ldots,N .
	\end{eqnarray}
	i.e.,
	%	\begin{eqnarray}\label{eq25}
	%		g_1 g_2/(\omega_{fg} - \omega_{1}) = g_1 g_2/(\omega_{fe} - \omega_{2}) \nonumber \\
	%		g_3 g_4/(\omega_{3} - \omega_{fg}) \nonumber\\
	%		g_3 g_4/(\omega_{4} - \omega_{fe})
	%	\end{eqnarray}
	
	\begin{eqnarray}\label{eq25a}
		\frac{g_1 g_2}{\Delta_{12}}
		=\frac{g_1 g_2}{\omega_{fg}-\omega_{1}}
		=\frac{g_1 g_2}{\omega_{fe}-\omega_{2}} .
	\end{eqnarray}
	\begin{eqnarray}\label{eq25b}
		\frac{g_{2j-1}g_{2j}}{\Delta_{(2j-1)(2j)}}
		=\frac{g_{2j-1}g_{2j}}{\omega_{\,2j-1}-\omega_{fg}}
		=\frac{g_{2j-1}g_{2j}}{\omega_{\,2j}-\omega_{fe}},
		\qquad j=2,3,\ldots,N .
	\end{eqnarray}
	than can be obtain by tunning the resonator frequencies, the coupling strengths, or qutrit energy level spacings.
	
From the above analysis, it is evident that throughout the W-state transfer
process the superconducting qutrit remains confined to the $\{\ket{g},\ket{e}\}$
manifold, while the upper energy level $\ket{f}$ is only virtually involved and
never populated. As a result, decoherence associated with relaxation and
dephasing of the $\ket{f}$ level is strongly suppressed, which significantly
enhances the robustness of the proposed hybrid CV-DV protocol.

Equation~(\ref{eq20}) shows that the transfer of the cat-state-encoded
$N$-qubit W state is implemented by a global unitary operator
$U = e^{-iH_0 \tau} e^{-iH_e \tau}$ acting on the initial state of the system.
Here, the transformation $e^{-iH_0 \tau}$ restores the system to the original
interaction picture, while the effective evolution governed by $H_e$ realizes
simultaneous beam-splitter-type interactions within all resonator pairs.

Consequently, the complete transfer of the hybrid CV-DV W state from the
odd-indexed resonators to the even-indexed resonators is achieved deterministically
in a single operational step, without the need for intermediate measurements,
feedforward control, or population of higher qutrit levels. This single-step
nature makes the protocol particularly well suited for scalable implementations
of $N$-qubit entanglement transfer in circuit QED architectures.
	
\section{Possible Experimental Implementation}

In the above analysis, we have considered a general type of microwave cavity, which can be either a three-dimensional (3D) cavity or a one-dimensional (1D) cavity. In this section, we discuss a feasible experimental implementation based on six transmission line resonators (TLRs) capacitively coupled to a single superconducting flux-type qutrit, as illustrated schematically in Fig.~2. Each TLR acts as a 1D microwave cavity.

For a superconducting flux device (e.g., a C-shunted flux qubit), the level spacings can be engineered to possess sufficiently large anharmonicity, and transitions between non-adjacent levels are allowed. This feature enables the use of the qutrit levels $|g\rangle$, $|e\rangle$, and $|f\rangle$, such that the resonator coupling with the $|g\rangle \leftrightarrow |f\rangle$ transition is accessible. In the following, we analyze the experimental feasibility of transferring multipartite entangled W states encoded in photonic cat-state qubits between spatially separated cavities.

When unwanted interactions are taken into account, the ideal Hamiltonian is modified as
\begin{equation}
	H' = H + \delta H_1 + \delta H_2 ,
\end{equation}
where $\delta H_1$ describes the unwanted inter-cavity crosstalk and $\delta H_2$ represents the pulse-induced unwanted transitions of the qutrit.

The inter-cavity crosstalk can be modeled as
\begin{equation}
	\delta H_1 = \sum_{j<l}^{6} g_{jl} \left( a_j a_l^\dagger e^{i\Delta_{jl} t} + \mathrm{H.c.} \right),
\end{equation}
where $g_{jl}$ is the crosstalk coupling strength between cavities $j$ and $l$, and $\Delta_{jl} = \omega_l - \omega_j$ is the frequency detuning between the two cavities. In the numerical simulations, we assume equal crosstalk strengths $g_{jl} \equiv g_{\mathrm{cr}}$ for simplicity.

The pulse-induced unwanted transition between the qutrit states $|e\rangle$ and $|f\rangle$ is described by
\begin{equation}
	\delta H_2 = \Omega_{fe} e^{i\Delta_p t} S_{fe}^{+} + \mathrm{H.c.},
\end{equation}
where $\Omega_{fe}$ is the Rabi frequency associated with the $|e\rangle \leftrightarrow |f\rangle$ transition, $\Delta_p = \omega_{fe} - \omega_{eg}$ is the detuning of the drive, and $S_{fe}^{+} = |f\rangle\langle e|$.

Due to the large anharmonicity of the flux qutrit, the pulse-induced $|g\rangle \leftrightarrow |f\rangle$ transition and cavity-induced transitions involving irrelevant levels are strongly suppressed. Therefore, their effects on the state-transfer performance are negligible and are not included in the numerical simulations.

By taking into account cavity dissipation as well as qutrit energy relaxation and dephasing, the system dynamics under the Markovian approximation is governed by the master equation
\begin{eqnarray}
	\frac{d\rho}{dt} &=& -i [H', \rho]
	+ \sum_{j=1}^{6} \kappa_j \mathcal{L}[a_j]
	+ \gamma_{eg} \mathcal{L}[\sigma^{-}_{eg}]
	+ \gamma_{fe} \mathcal{L}[\sigma^{-}_{fe}]
	+ \gamma_{fg} \mathcal{L}[\sigma^{-}_{fg}] \nonumber \\
	& & + \sum_{l=e,f} \gamma_{\phi,l}
	\left(
	\sigma_{ll} \rho \sigma_{ll}
	- \frac{1}{2} \sigma_{ll} \rho
	- \frac{1}{2} \rho \sigma_{ll}
	\right),
\end{eqnarray}
where $\mathcal{L}[\Lambda] = \Lambda \rho \Lambda^\dagger - \Lambda^\dagger \Lambda \rho /2 - \rho \Lambda^\dagger \Lambda /2$.
Here, $\kappa_j$ is the decay rate of cavity $j$, $\gamma_{eg}$ is the relaxation rate for the $|e\rangle \rightarrow |g\rangle$ transition, $\gamma_{fe}$ ($\gamma_{fg}$) denotes the relaxation rate for the $|f\rangle \rightarrow |e\rangle$ ($|f\rangle \rightarrow |g\rangle$) transition, and $\gamma_{\phi,e}$ ($\gamma_{\phi,f}$) is the pure dephasing rate of the level $|e\rangle$ ($|f\rangle$).

The fidelity of the entangled state transfer is defined as
\begin{equation}
	F = \sqrt{\langle \psi_{\mathrm{id}} | \rho_{\mathrm{cav}} | \psi_{\mathrm{id}} \rangle},
\end{equation}
where $|\psi_{\mathrm{id}}\rangle$ is the ideal target W-type cat-state of the six cavities, and $\rho_{\mathrm{cav}}$ is the reduced density operator of the cavity subsystem obtained by tracing out the qutrit degrees of freedom. The operation time is chosen as $t = T = \pi / (2|\lambda_{12}|)$.

\begin{figure}[t]
		\centering
	\includegraphics[width=12cm, height=8cm]{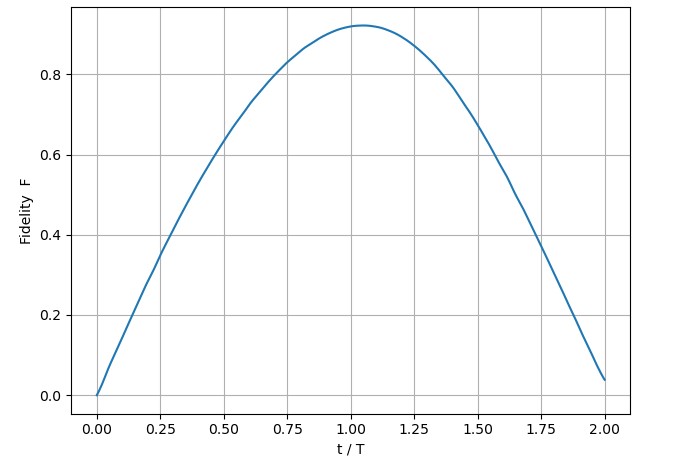}
	\caption{Time evolution of the fidelity $F$ for the transfer of the W-type cat-state encoded cavity state as a function of the normalized interaction time $t/T$, where $T=\pi/(2|\lambda_{12}|)$ is the effective swap time. The simulation includes cavity dissipation ($\kappa^{-1}=20~\mu\mathrm{s}$), qutrit energy relaxation and dephasing, inter-cavity crosstalk, and unwanted pulse-induced transitions using experimentally realistic parameters.}
	\label{fig:fidelity_tT}
\end{figure}

For a superconducting flux qutrit, the transition frequencies between adjacent levels can be tuned within the range of $5$--$20~\mathrm{GHz}$. In the simulations, we choose detunings $\Delta_j/2\pi = 0.40$--$0.45~\mathrm{GHz}$ and coupling strengths $g_j/2\pi = 150$--$175~\mathrm{MHz}$, which are well within experimentally demonstrated values.

The cavity frequencies are chosen in the range $4.2$--$14.2~\mathrm{GHz}$, yielding large inter-cavity detunings that suppress unwanted resonant interactions. The unwanted $|e\rangle \leftrightarrow |f\rangle$ drive is taken as $\Omega_{fe}/2\pi = 47~\mathrm{MHz}$ with detuning $\Delta_p/2\pi = -2.5~\mathrm{GHz}$.

The cavity photon lifetime is chosen as $\kappa^{-1} = 20~\mu\mathrm{s}$, corresponding to quality factors on the order of $10^6$. The qutrit coherence parameters are conservatively taken as
\begin{eqnarray}
	T_{1}^{eg} &=& 30~\mu\mathrm{s}, \quad
	T_{1}^{fe} = 20~\mu\mathrm{s}, \quad
	T_{1}^{fg} = 60~\mu\mathrm{s}, \nonumber \\
	T_{\phi}^{e} &=& 40~\mu\mathrm{s}, \quad
	T_{\phi}^{f} = 25~\mu\mathrm{s}.
\end{eqnarray}
The cat-state amplitude is chosen as $\alpha = 0.5$.

Numerical simulations show that the effect of inter-cavity crosstalk is weak even when the crosstalk strength reaches a few percent of the effective cavity--cavity coupling. Moreover, the decay of the second excited state of the qutrit has a negligible effect on the operation fidelity due to the short interaction time compared with the coherence times. Using the experimentally realistic parameters considered above, a maximum fidelity of about $0.92$ is obtained at the optimal interaction time as shown in figure \ref{fig:fidelity_tT}. These results demonstrate that the proposed one-step transfer of W states encoded in photonic cat-state qubits is experimentally feasible with current circuit QED technology.
	
\section{Conclusion}

In this work, we have proposed a deterministic and single-step protocol for transferring
multipartite W-type entanglement encoded in photonic Schr\"odinger cat-state qubits within
a circuit QED architecture. By exploiting dispersive Raman-type interactions mediated by a
single superconducting flux qutrit, the scheme enables simultaneous beam-splitter--type
couplings between pairs of microwave resonators, allowing the coherent transfer of an
$n$-qubit hybrid continuous-variable--discrete-variable (CV--DV) W state without populating
higher excited atomic levels.

A central advantage of the protocol is that the superconducting qutrit remains effectively
confined to its lower two energy levels throughout the evolution, while the highest excited
state is only virtually involved. This significantly suppresses decoherence arising from
energy relaxation and dephasing and enhances the robustness of the transfer process.
Moreover, the entire operation is completed in a single collective step and does not rely
on intermediate measurements, feedforward control, or sequential gate operations, thereby
reducing operational complexity and improving scalability.

We have assessed the performance of the scheme by numerically solving the full Lindblad
master equation, incorporating realistic sources of decoherence including cavity photon
loss, qutrit energy relaxation and dephasing, inter-cavity crosstalk, and pulse-induced
unwanted transitions. For experimentally accessible parameters in superconducting circuit
platforms, our simulations demonstrate that the transfer of a three-qubit cat-state W state
can be achieved with a maximum fidelity of approximately $0.92$. These results confirm that
the proposed hybrid CV--DV W-state transfer protocol is feasible with current circuit QED
technology and provides a practical route toward scalable multipartite entanglement
distribution in superconducting microwave quantum networks.

	\section*{References}
	\bibliographystyle{iopart-num}
	\bibliography{refr}

\end{document}